\documentclass[aoas,nameyear,secthm,seceqn,dvips]{arximspdf}
\usepackage{dcolumn}
\usepackage{graphicx}

\doi{10.1214/10-AOAS331}
\volume{4}
\issue{3}
\pubyear{2010}
\firstpage{1383}
\lastpage{1402}

\makeatletter
\newcolumntype{d}[1]{D{.}{.}{#1}}
\makeatother

\begin{document}
\begin{frontmatter}

\title{Poisson point process models solve\\ the ``pseudo-absence
problem'' for\\ presence-only data in ecology\thanksref{TITL1}}%\protect{T1}}
\pdftitle{Poisson point process models solve the ``pseudo-absence
problem'' for presence-only data in ecology}
\runtitle{Point process models for species distribution modeling}
\thankstext{TITL1}{Supported by the Australian Research Council, Linkage Project LP0774833.}

\begin{aug}
\author[A]{\fnms{David I.} \snm{Warton}\corref{}\ead[label=e1]{David.Warton@unsw.edu.au}}
and
\author[B]{\fnms{Leah C.} \snm{Shepherd}\ead[label=e2]{leah.shepherd@iinet.net.au}}

\runauthor{D. I. Warton and L. C. Shepherd}

\affiliation{University of New South Wales}
\address[A]{School of Mathematics and Statistics\\
\quad and Evolution \& Ecology Research Centre\\
 University of New South Wales\\
Sydney, NSW 2052\\
Australia\\
\printead{e1}} %adresu isvedimo komanda gale!
\address[B]{School of Mathematics and Statistics\\
 University of New South Wales\\
Sydney, NSW 2052\\
Australia\\
\printead{e2}}
\end{aug}

% HISTORY:
\received{\smonth{5} \syear{2009}}
\revised{\smonth{12} \syear{2009}}

% ABSTRACT
%
\begin{abstract}
Presence-only data, point locations where a species has been recorded
as being present, are often used in modeling the distribution of a
species as a function of a set of explanatory variables---whether to
map species occurrence, to understand its association with the
environment, or to predict its response to environmental change.
Currently, ecologists most commonly analyze presence-only data by
adding randomly chosen ``pseudo-absences'' to the data such that it can
be analyzed using logistic regression, an approach which has weaknesses
in model specification, in interpretation, and in implementation. To
address these issues, we propose Poisson point process modeling of the
intensity of presences. We also derive a link between the proposed
approach and logistic regression---specifically, we show that as the
number of pseudo-absences increases (in a regular or uniform random
arrangement), logistic regression slope parameters and their standard
errors converge to those of the corresponding Poisson point process
model. We discuss the practical implications of these results. In
particular, point process modeling offers a framework for choice of the
number and location of pseudo-absences, both of which are currently
chosen by ad hoc and sometimes ineffective methods in ecology, a
point which we illustrate by example.
\end{abstract}

% KEYWORDS


\end{frontmatter}

%s1 ###
\section{Background}
\label{s:back}
\begin{figure}

\includegraphics{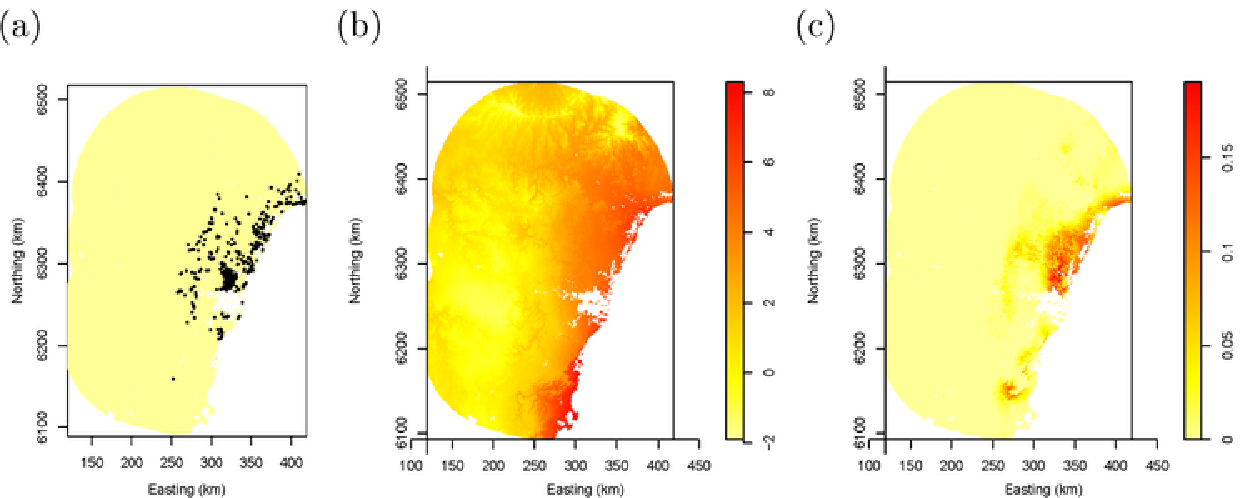}

\caption{\textup{(a)} Example presence-only data---atlas records of where the
tree species \textup{Angophora costata} has been reported to be present,
west of Sydney, Australia. The study region is shaded. \textup{(b)} A map of
minimum temperature \textup{($^{\circ}$C)} over the study region. Variables such as
this are used to model how intensity of \textup{A.\ costata} presence
relates to the environment. \textup{(c)} A species distribution model, modeling
the association between \textup{A.\ costata} and a suite of environmental
variables. This is the fitted intensity function for \textup{A.\ costata}
records per km$^2$, modeled as a quadratic function of four
environmental variables using a point process model as in Section~\protect\ref{s:data}.
}\label{f:angoPoints}
\end{figure}

Pearce and Boyce (\citeyear{PearceBoyce}) define presence-only data as ``consisting only of
observations of the organism but with no reliable data
on where the species was not found. Sources for these
data include atlases, museum and herbarium records,
species lists, incidental observation databases and
radio-tracking studies.'' Note that such data arise as \textit{point
locations} where the organism is observed, which we denote as $\mathbf
{y}$ in this article. An example is given in Figure~\ref
{f:angoPoints}(a). This figure gives all locations where a particular
tree species (\textit{Angophora costata}) has been reported by park
rangers since 1972, within 100 km of the Greater Blue Mountains World
Heritage Area, near Sydney, Australia. Note that this does not consist
of all locations where an \textit{Angophora costata} tree is found---rather it is the locations where the species has been \textit{reported}
to be found. We would like to use these presence points, together with
maps of explanatory variables describing the environment (often
referred to in ecology as ``environmental variables''), to predict the
location of \textit{A.\ costata} and how it varies as a function of
explanatory variables (Figure~\ref{f:angoPoints}).

%f1 ###

Presence-only data are used extensively in ecology to model species
distribu\-tions---while the term ``presence-only data'' was rarely used
before the 1990s, ISI Web of Science reports that it was used in 343
publications from 2005 to 2008. The use of presence-only data in
modeling is a relatively recent development, presumably aided by the
movement toward electronic record keeping and recent advances in
Geographic Information Systems. One reason for the current widespread
usage of presence-only data is that often this is the best available
information concerning the distribution of a species, as there is often
little or no information on species distribution being available from
systematic surveys [\citet{ElithLeathwick2007}].

Species distribution models, sometimes referred to as habitat models or
habitat classification models [\citet{Zarnetskeetal}], are regression
models for the likelihood that a species is present at a given
location, as a function of explanatory variables that are available
over the whole study region. Such models are used to construct maps
predicting the full spatial distribution of a species [given GIS maps
of explanatory variables such as in Figure~\ref{f:angoPoints}(b)]. When
surveys have recorded the presence and absence of a species in a
pre-defined study area (``presence/absence data''), logistic regression
approaches and modern generalizations [\citet{ElithBRT}] are typically
used for species distribution modeling. If instead presence-only data
are to be used in species distribution modeling, then a common approach
to analysis is to first create ``pseudo-absences,'' denoted as $\mathbf
{y}_0$, usually achieved by randomly choosing point locations in the
region of interest and treating them as absences. Then the
presence/pseudo-absence data set is analyzed using standard analysis
methods for presence/absence data [\citet{PearceBoyce}; \citet{ElithLeathwick2007}], which have been used in species
distribution modeling for a long time [\citet{Austin85}].
\citet{Wardetal09} recently proposed a modification of the
pseudo-absence logistic regression approach for the analysis of
presence-only data, when the probability $\pi$ that a randomly chosen
pseudo-absence point of a presence is known. However, $\pi$ is not
known in practice.

We see three key weaknesses of the ``pseudo-absence'' approach so
widely used in ecology for analyzing presence-only data, which we
describe concisely as problems of model specification, interpretation,
and implementation. A sounder \textit{model specification} would involve
constructing a model for the observed data $\mathbf{y}$ \textit{only},
rather than requiring us to generate new data $\mathbf{y}_0$ prior to
constructing a model. \textit{Interpretation} of results is difficult,
because some model parameters of interest (such as $p_i$ of Section~\ref
{s:pseudo}) are a function of the number of pseudo-absences and their
location. For example, we explain in Section~\ref{s:pseudo} that as the
number of pseudo-absences approaches infinity, $p_i\rightarrow0$, for
a given presence-only data set $\mathbf{y}$. \textit{Implementation} of
the approach is problematic because it is unclear how pseudo-absences
should be chosen [\citet{ElithLeathwick2007}; \citet{Guisanetal2007Grain}; \citet{Zarnetskeetal}; \citet{Phillipsetal09}],
and one can obtain qualitatively different results depending on the
method of choice of pseudo-absences [\citet{ChefaouiLobo}].

In this paper we make two key contributions. First, we propose point
process models (Section~\ref{s:ppm}) as an appropriate tool for species
distribution modeling of presence-only data, given that presence-only
data arise as a set of point events---a set of locations where a
species has been reported to have been seen. A point process model
specification addresses each of the three concerns raised above
regarding pseudo-absence approaches. Our second key contribution is a
proof that the pseudo-absence logistic regression approach, when
applied with an increasing number of regularly spaced or randomly
chosen pseudo-absences, yields estimates of slope parameters that
converge to the point process slope estimates (Section~\ref{s:pseudo}).
These two key results have important ramifications for species
distribution modeling in ecology (Section~\ref{s:discuss}), in
particular, we provide a solution to the problem of how to select
pseudo-absences. We illustrate our results for the \textit{A.\ costata}
data of Figure~\ref{f:angoPoints}(a) (Section~\ref{s:data}).

%s2 ###
\section{Poisson point process models for presence-only data}\label{s:ppm}

Presence-only data are a set $\mathbf{y}=\{ y_1,\ldots,y_n \}$ of point
locations in a two-dimensional region $\mathcal{A}$, where the
locations where presences are recorded (the $y_i$) are out of the
control of the researcher, as is the total number of presence points
$n$. We also observe a ``map'' of values over the entire region
$\mathcal{A}$ for each of $k$ explanatory variables, and we denote the
values of these variables at $y_i$ as $(x_{i1},\ldots,x_{ik})$.

We propose analyzing $\mathbf{y}=\{ y_1,\ldots,y_n \}$ as a point
process, hence, we jointly model number of presence points $n$ and
their location ($y_i$). This has not previously been proposed for the
analysis of presence-only data, despite the extensive literature on the
analysis of presence-only data. We consider inhomogeneous Poisson point
process models [\citet{CressieBook}; \citet{DiggleBook}], which make the
following two assumptions:
\begin{enumerate}
\item The locations of the $n$ point events $(y_1,\ldots,y_n)$ are independent.
\item The intensity at point $y_i$ [$\lambda(y_i)$, denoted as $\lambda
_i$ for convenience], the limiting expected number of presences per
unit area [\citet{CressieBook}], can be modeled as a function of the $k$
explanatory variables. We assume a log-linear specification:
%
%e2.1 ###
\begin{equation}\label{e:ppmModel}
\log(\lambda_i) = \beta_0 + \sum_{j=1}^k x_{ij}\beta_j,
\end{equation}
although note that the linearity assumption can be relaxed in the usual
way (e.g., using quadratic terms or splines). The parameters of
the model for the $\lambda_i$ are stored in the vector $\beta=(\beta_0,
\beta_1,\ldots,\beta_k)$.
\end{enumerate}
Note that the process being modeled here is locations where an organism
\textit{has been reported} rather than locations where individuals of the
organism occur. Hence, the independence assumption would only be
violated by interactions between records of sightings rather than by
interactions between individual organisms per se. The atlas data
of Figure~\ref{f:angoPoints} consist of 721 \textit{A.\ costata} records
accumulated over a period of 35 years in a region of 86,000 km$^2$, so
independence of records seems a reasonable assumption in this case,
given the rarity of event reporting. Nevertheless, the methods we
review here can be generalized to handle dependence between point
events [\citet{Spatstat}].

\citet{CressieBook} shows that the log-likelihood for $\mathbf{y}$ can
be written as
%
%e2.2 ###
\begin{equation}\label{e:ppmint}
l(\beta;\mathbf{y}) = \sum_{i=1}^n \log(\lambda_i) - \int_{y\in\mathcal
{A}} \lambda(y) \,dy - \log(n!).
\end{equation}
\citet{BermanTurner} showed that if the integral is estimated via
numerical quadrature as $\int_{y\in\mathcal{A}} \lambda(y) \,dy \approx
\sum_{i=1}^m w_i \lambda_i$, then the log-likelihood is (approximately)
proportional to a weighted Poisson likelihood:
%
%e2.3 ###
\begin{equation}\label{e:ppmLL}
l_{\mathrm{ppm}}(\beta;\mathbf{y},\mathbf{y}_0,\mathbf{w}) = \sum_{i=1}^m
w_i \bigl( z_i\log(\lambda_i) - \lambda_i \bigr),
\end{equation}
where $z_i=\frac{I(i\in\{1,\ldots,n \})}{w_i}$, $\mathbf{y}_0=\{
y_{n+1},\ldots,y_m\}$ are quadrature points, the vector $\mathbf{w}=(
w_1,\ldots, w_m )$ stores all quadrature weights, and $I(\cdot)$ is the
indicator function. Being able to write $l(\beta;\mathbf{y})$ as a
weighted Poisson likelihood has important practical significance
because it implies that generalized linear modeling (GLM) techniques
can be used for estimation and inference about $\beta$. Further,
adaptations of GLM techniques to other settings, such as generalized
additive models [\citet{GAMbook}], can then be readily applied to
Poisson point process models also.

Before implementing this approach, however, we need to make two key
decisions---how to choose quadrature points $\mathbf{y}_0=\{ y_{n+1},
\ldots, y_m \}$ and how to calculate the quadrature weight $w_i$ at
each point $y_i$.

We propose choosing quadrature points in a regular rectangular grid,
and considering grids of increasing spatial resolution until the
estimate of the maximized log-likelihood
$l_{\mathrm{ppm}}(\hat{\beta};\mathbf{y},\mathbf{y}_0,\mathbf{w})$
has converged. A rectangular grid provides reasonably efficient
coverage of the region $\mathcal{A}$, and is an arrangement for which
environmental data $x_{i1},\ldots,x_{ik}$ can be easily obtained via
GIS software. We illustrate this method in Section~\ref{s:data}. Note a
large data set may be required---in Section~\ref{s:data} convergence
was achieved at a spatial scale that required inclusion of
approximately 86,000 quadrature points.

Quadrature weights are calculated as the area of the neighborhood $A_i$
around each point $y_i$, according to some definition of the $A_i$ such
that $y_i\in A_i$ for each $i$, $A_i \cap A_{i'} = \varnothing$ for
each $i\neq i'$, and $\bigcup_i A_i = \mathcal{A}$. In Section~\ref
{s:data} we calculated quadrature weights using the tiling method
implemented in the \texttt{R} package \texttt{spatstat} [\citet{Spatstat}]. This
crude approach breaks the region $\mathcal{A}$ into rectangular tiles
and calculates the weight of a point as the inverse of the number of
points per unit area in its tile. We fixed tile size at the size of the
regular grid used to sample quadrature points, such that all tiles
contained exactly one quadrature point. Dirichlet tessellation [\citet
{Spatstat}] offers an alternative method of estimating weights, but
this was not practical for our sample sizes.

%s3 ###
\section{Asymptotic equivalence of pseudo-absence logistic regression
and Poisson point process models}\label{s:pseudo}
Ecologists typically analyze presence-only data points $\mathbf{y} = \{
y_{1}, \ldots, y_n \}$ by generating a set of ``pseudo-absence'' points
$\mathbf{y}_0 = \{ y_{n+1}, \ldots, y_m \}$, then using logistic
regression to model the ``response variable'' $I(i\in\{1,\ldots, n\})$
as a function of explanatory variables [\citet{PearceBoyce}]. Note that
$I(i\in\{1,\ldots, n\})$ is not actually a stochastic quantity,
nevertheless, the use of logistic regression to model this quantity as
a Bernoulli response variable can be motivated via a case-control
argument along the lines of \citet{DiggleBook}, Section 9.3.

In this section we will show that the approach to analysis currently
used in ecology, logistic regression using pseudo-absences, is closely
related to the Poisson point process model introduced in Section~\ref
{s:ppm}. Specifically, if the pseudo-absences are either generated on a
regular grid or completely at random over the region $\mathcal{A}$,
then as the number of pseudo-absences increase, all parameter
estimators except for the intercept in the logistic regression model
converge to the maximum likelihood estimators of the Poisson process
model of Section~\ref{s:ppm}. This asymptotic relationship between
logistic regression and Poisson point process models does not appear to
have been recognized previously in the literature.

First, we will specify a probability model for $I(i\in\{1,\ldots, n\}
)$ that permits a logistic regression model, and the study of its
properties as $m\rightarrow\infty$.
This can be achieved by considering a point chosen at random from $\{
y_{1}, \ldots, y_m \}$ and defining $U$ as the event that the randomly
chosen point $y_i$ is a presence. We are interested in modeling $U$
conditionally on the explanatory variables observed at the randomly
chosen point, $\mathbf{x}_i=(x_{i1},\ldots,x_{ik}$). In this setting, $U$
is a Bernoulli variable with conditional mean $p_i$, and we assume that
%
%e3.1 ###
\begin{equation}\label{e:binModel}
\log \biggl( \frac{p_i}{1-p_i} \biggr) = \gamma_0-\log(m-n) + \sum_{j=1}^k
x_{ij}\gamma_j.
\end{equation}
The intercept term is written as $\gamma_0-\log(m-n)$ because
%
%e3.2 ###
\begin{equation}\label{e:Bayes}
\frac{p_i}{1-p_i} = \frac{f_1(\mathbf{x}_i|U=1)}{f_0(\mathbf
{x}_i|U=0)}\cdot\frac{P(U=1)}{P(U=0)} = \frac{f_1(\mathbf
{x}_i|U=1)}{f_0(\mathbf{x}_i|U=0)} \frac{n}{m-n},
\end{equation}
where $f_1(\cdot)$ and $f_0(\cdot)$ are the densities of $\mathbf{x}_i$
conditional on $U=1$ and $U=0$ respectively. Provided that $f_0(\mathbf
{x}_i|U=0)$ is not a function of $m$ (which is ensured, e.g., by
using an identical process to select all pseudo-absence points), then
the odds of a presence point $\frac{p_i}{1-p_i}$ is a function of $m$
only via the multiplier $(m-n)^{-1}$.

It can be seen from equation~(\ref{e:Bayes}) that if $m\rightarrow\infty$
in such a way that $f_0(\mathbf{x}_i|U=0)$ is not a function of $i$,
then $p_i\rightarrow0$ at an asymptotic rate that is proportional to
$m^{-1}$, and the intercept term in the logistic regression model
approaches $-\infty$ at the rate $\log(m)$. This in turn means that the
logistic regression log-likelihood, defined below, will also diverge as
$m\rightarrow\infty$:
%
%e3.3 ###
\begin{equation}\label{e:binll}
l_{\mathrm{bin}}(\gamma; \mathbf{y}, \mathbf{y}_0 ) = \sum_{i=1}^n \log
(p_i) + \sum_{i=n+1}^m \log(1-p_i).
\end{equation}
Clearly, as $p_i\rightarrow0$, $\log(p_i)\rightarrow-\infty$ and,
hence, $l_{\mathrm{bin}}(\gamma; \mathbf{y}, \mathbf{y}_0 ) \rightarrow
-\infty$. Such divergence is a symptom that the original model has been
incorrectly specified. The use of the more appropriate spatial point
process model of Section~\ref{s:ppm} will not encounter such problems.
However, it is shown in the following theorems that despite the
problems inherent in the logistic regression model specification, and
despite divergence of the intercept term, the remaining parameters
converge to the corresponding parameters from the Poisson process model
of equation~(\ref{e:ppmLL}). Further, pseudo-absences play the same role
in the logistic regression model that quadrature points played in
Section~\ref{s:ppm}.

For notational convenience, we will define $J_m$ to be the single-entry
matrix whose first element is $\log m$:
%
%e3.4 ###
\begin{equation}\label{e:J}
J_m = (\log m, 0, \ldots, 0 ).
\end{equation}
This definition will be used in each of the theorems that follow. The
$J_m$ notation is immediately useful in writing out the parameters of
the model for $p_i$ in equation~(\ref{e:binModel}) as $\gamma- J_{m-n}$
where $\gamma= (\gamma_0,\gamma_1,\ldots,\gamma_k)$.

\begin{thm}\label{t:ll}
Consider a fixed set of $n$ observations from a point process $\mathbf
{y}=\{ y_1,\ldots,y_n \}$, and a set of pseudo-absences $\mathbf{y}_0 =
\{ y_{n+1}, \ldots, y_m \}$ of variable size that is chosen via some
identical process on $\mathcal{A}$ for $i\in\{n+1,\ldots,m\}$. We model
$U$, whether or not a randomly chosen point is a presence point, via
logistic regression as in equation~(\ref{e:binModel}).

As $m\rightarrow\infty$, the logistic regression log-likelihood of
equation~(\ref{e:binll}) approaches the Poisson point process
log-likelihood [equation~(\ref{e:ppmLL})] but with all quadrature weights
set to one:
\[
l_{\mathrm{bin}}(\gamma; \mathbf{y}, \mathbf{y}_0 ) = l_{\mathrm{ppm}}(\gamma
-J_m; \mathbf{y}, \mathbf{y}_0, \mathbf{1}) + O(m^{-1}),
\]
where $\mathbf{1}$ is a $m$-vector of ones, and $J_m$ is defined in
equation~(\ref{e:J}).
\end{thm}

The proofs to Theorem~\ref{t:ll} and all other theorems are given in
Appendix~\ref{s:appendix}.

Theorem~\ref{t:ll} has two interesting practical implications.

\begin{figure}

\includegraphics{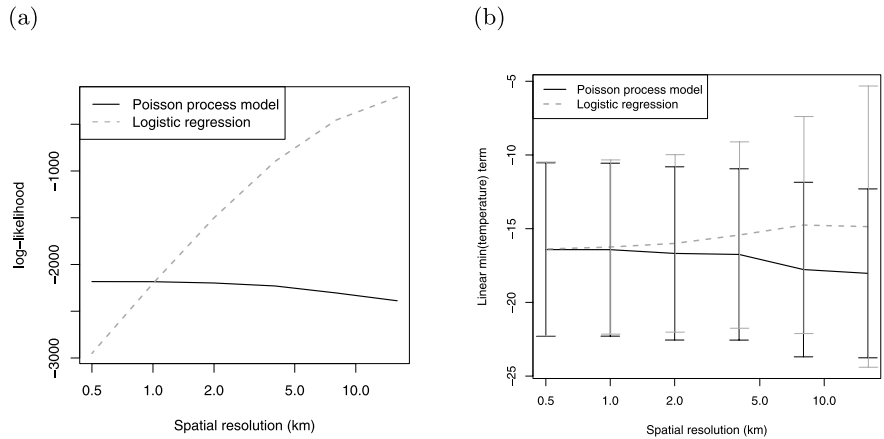}

\caption{Asymptotic behavior of Poisson point process and
pseudo-absence logistic regression models when the number of quadrature
points becomes large (via sampling in a regular grid with increasing
spatial resolution). \textup{(a)} The maximized log-likelihood converges for a
Poisson point process, but not for pseudo-absence logistic regression.
\textup{(b)} The parameters and their standard errors converge for Poisson point
process and logistic regression models, for sufficiently high spatial
resolution. Linear coefficient of ``minimum temperature'' is given here
(corresponding to the second entry in Table~\protect\ref{tab:params}).}\label{f:res}
\end{figure}

First, it implies that the pseudo-absence points of presence-only
logistic regression play the same role as quadrature points of a point
process model, and so established guidelines on how to choose
quadrature points (such as those of Section~\ref{s:ppm}) can inform
choice of pseudo-absences. Previously pseudo-absences have been
generated according to ad hoc recommendations [\citet{PearceBoyce}; \citet{Zarnetskeetal}], given the lack of a theoretical framework
for their selection. In contrast, quadrature points are generated in
order to estimate the log-likelihood to a pre-determined level of
accuracy, a criterion which guides the choice of locations and numbers
of quadrature points $m-n$, as explained in Section~\ref{s:ppm} and as
illustrated later in Section~\ref{s:data} [Figure~\ref{f:res}(a)].
Interestingly, current methods of selecting pseudo-absences in ecology
[\citet{PearceBoyce}; \citet{Zarnetskeetal}] do not appear to be consistent with
the best practice in low-dimensional numerical quadrature---points are
usually selected at random rather than on a regular grid, and the
number of pseudo-absences ($m-n$) is more commonly chosen relative to
the magnitude of the number of presences ($n$) rather than based on
some convergence criterion as in Figure~\ref{f:res}(a).

%f2 ###

Second, Theorem~\ref{t:ll} implies that despite the apparent ad
hoc nature of the pseudo-absence approach, \textit{some form of point
process model is being estimated}. However, logistic regression is only
equivalent to a Poisson point process when $\mathbf{w}=\mathbf{1}$,
that is, all quadrature weights are ignored. The implications of
ignoring weights is considered in Theorems~\ref{t:weights} and~\ref
{t:random} below.

It should also be noted that Theorem~\ref{t:ll} is closely related to
results due to \citet{Owen07} and \citet{WardPhD}, although Theorem~\ref
{t:ll} differs from these results by relating pseudo-absence logistic
regression specifically to point process modeling.

\citet{Owen07} also considered the logistic regression setting where
the number of presence points is fixed, and the number of
pseudo-absences increases to infinity, and referred to this as
``infinitely unbalanced logistic regression.'' \citet{Owen07} derived
conditions under which convergence of model parameters could be
achieved as the number of pseudo-absences increased. The key condition
is that the centroid of the points $\{ y_1,\ldots,y_n \}$ in the design
space is ``surrounded''---see Definition 3 of \citet{Owen07} for details.

Along the lines of \citet{Wardetal09}, \citet{WardPhD} considered a
pseudo-absence logistic regression formulation of the presence-only
data problem, and defined the ``population logistic model'' as the
model across ``the full population'' of locations in the region
$\mathcal{A}$. The unconstrained log-likelihood of the population
logistic model [\citet{WardPhD}, equation (7.6)] has a similar form to the
point process log-likelihood $l(\beta;\mathbf{y})$ of Section~\ref
{s:ppm}. For presence-only logistic regression as in \citet
{Wardetal09}, \citet{WardPhD} shows that as the number of
pseudo-absences approaches infinity, the log-likelihood converges to
that of the population logistic model, a result that is analogous to
Theorem~\ref{t:ll}.

Having shown that pseudo-absence logistic regression is equivalent to a
point process model where weights are ignored, the implications of
ignoring weights is now considered in Theorems~\ref{t:weights} and~\ref
{t:random}.
\begin{thm}\label{t:weights}
Consider a point process model with quadrature points $y_{n+1},\ldots
,y_m$ selected such that for all $i$ $w_i=\frac{|\mathcal{A}|}{m}$,
where $|\mathcal{A}|$ is the total area of the region $\mathcal{A}$.
Assume also that the design matrix $\mathbf{X}$ has full rank.

The maximum likelihood estimators of $l_{\mathrm{ppm}}(\gamma- J_m;\mathbf
{y},\mathbf{y}_0,\mathbf{1})$ and $l_{\mathrm{ppm}}(\beta;\mathbf{y},\break\mathbf
{y}_0,\frac{|\mathcal{A}|}{m}\mathbf{1})$, $\hat{\gamma} - J_m$ and
$\hat{\beta}$ respectively, satisfy
\[
\hat{\gamma} = \hat{\beta} + J_{|\mathcal{A}|}.
\]
Further, the Fisher information for $\hat{\gamma}$ and $\hat{\beta}$ is equal.

That is, provided that quadrature points have been selected such that
quadrature weights are equal, ignoring quadrature weights in a Poisson
point process model does not change slope parameters nor their standard
errors, although the intercept term will differ by $\log(|\mathcal{A}|/m)$.
\end{thm}

Theorem~\ref{t:weights} refers to the special case where all points
(including presence points) are sampled on a regular grid. This arises
in the special case where the region $\mathcal{A}$ has been divided
into grid cells of equal area, and each grid cell is assigned the value
1 only if it contains a presence point. This form of presence-only
analysis is sometimes used in ecology [\citet{MAXENT2}, e.g.]. In
addition, the setting of Theorem~\ref{t:weights} provides a reasonable
approximation to the approach to quadrature-point selection proposed in
Section~\ref{s:ppm}.

Note that together Theorems~\ref{t:ll}--\ref{t:weights} suggest that
when quadrature points (or, equivalently, pseudo-absences) are sampled
in a regular grid at increasing resolution, the logistic regression
parameter estimates and their standard errors will approach those of
the point process model---with the exception of the intercept term,
which diverges slowly to $-\infty$ as all $p_i\rightarrow0$ at a rate
inversely proportional to $m$. This nonconvergence of the intercept was
also noticed by \citet{Owen07}.
Figure~\ref{f:res} illustrates these results for the \textit{A.\
costata} data.

Theorem~\ref{t:random} below links the above results with the case
where pseudo-absences are randomly sampled within the region $\mathcal
{A}$, which is a more common approach in ecology than sampling on a
regular grid [e.g., \citet{ElithLeathwick2007}; \citet{Hernandezetal}].

\begin{thm}\label{t:random}
Consider again the conditions of Theorem~\ref{t:weights}, but now
assume that the quadrature points $\mathbf{y}_0$ are selected uniformly
at random within the region $\mathcal{A}$. As previously, $\hat{\gamma
}$ is the maximum likelihood estimator of $l_{\mathrm{ppm}}(\gamma
-J_m;\mathbf{y},\mathbf{y}_0,\mathbf{1})$, but now let $\hat{\beta}$ be
the maximum likelihood estimator of $l(\beta;\mathbf{y})$ from
equation~(\ref{e:ppmint}). As $m\rightarrow\infty$,
\[
\hat{\gamma} \stackrel{\mathcal{P}}{\rightarrow} \hat{\beta} +
J_{|\mathcal{A}|}.
\]
That is, if quadrature points are randomly selected instead of being
sampled on a regular grid, the result of Theorem~\ref{t:weights} holds
in probability rather than exactly.
\end{thm}

Note that the stochastic convergence in Theorem~\ref{t:random} is with
respect to $m$ not $n$, that is, it is conditional on the observed
point process.

Note also that one can think of randomly selecting pseudo-absence
points as an implementation of ``crude'' Monte Carlo integration [\citet
{Lepage1978}] for estimating $\int_{y\in\mathcal{A}} \lambda(y) \,dy$ in
the point process likelihood [equation~(\ref{e:ppmint})].

%s4 ###
\section{Modeling \textit{Angophora costata} species distribution}\label{s:data}

As an illustration, we construct Poisson point process models for the
intensity of \textit{Angophora costata} records as a function of a set of
explanatory variables. We consider modeling the log of intensity using
linear and quadratic functions of the variables minimum and maximum
temperature, mean annual rainfall, number of fires since 1943, and
``wetness,'' a coefficient which can be considered as an indicator of
local moisture. These five variables were recommended by local experts
as likely to be important in determining \textit{A.\ costata} distribution.

Our full model for intensity of \textit{A.\ costata} records at the point
$y_i$ has the following form:
%
%e4.1 ###
\begin{equation} \label{e:model}
\log(\lambda_i) = \beta_0 + \mathbf{x}_i^T \beta_1 + \mathbf{x}_i^T
\mathbf{B} \mathbf{x}_i,
\end{equation}
where $\beta_1$ is a vector of linear coefficients, $\mathbf{B}$ is a
matrix of quadratic coefficients, and $\mathbf{x}_i$ is a vector
containing measurements of the five environmental variables at point
$y_i$. We consider a quadratic model for $\log(\lambda_i)$ because this
enables fitting a nonlinear function and interaction between different
environmental variables, both considered important in species
distribution modeling [\citet{ElithBRT}].

All analyses were carried out using purpose-written code on the \texttt
{R} program [\citet{Rpackage}].

We first considered the spatial resolution at which quadrature points
needed to be sampled in order for the log-likelihood $l(\beta; \mathbf
{y})$ to be suitably well approximated by $l_{\mathrm{ppm}}(\beta; \mathbf
{y}, \mathbf{y}_0, \mathbf{w})$. We found [as in Figure~\ref{f:res}(a)]
that on increasing the number of quadrature points, the estimate of the
maximized log-likelihood converged, and that there was minimal change
in the solution beyond a resolution of one quadrature point every 1 km
(the maximized log-likelihood changed by less than one when the number
of quadrature points was increased 4-fold). Hence, the 1 km resolution
was used in model-fitting, and these results are reported here. This
involved a total of 86,227 quadrature points.

In order to study which environmental variables are associated with
\textit{A.\ costata} and how they are associated, we performed model
selection where we considered different forms of models for
log-intensity as a function of environmental variables, and we
considered different subsets of the environmental variables via
all-subsets selection. In both cases we used AIC as our model selection
criterion, a simple and widely-used penalty-based model selection
criterion [\citet{BurnhamAnderson}].

Comparison of AIC values for linear and quadratic models suggest that a
much better fit is achieved when using a quadratic model with
interactions terms for all coefficients (Table~\ref{tab:AIC}). Hence,
we have evidence that environmental variables interact in their effect
on \textit{A.\ costata}. Judging from the model coefficients and their
relative size compared to standard errors, the interactions between
maximum temperature, minimum temperature, and annual rainfall appear to
be the major contributors.

%t1 ###
\begin{table}
\tablewidth=275pt
\caption{AIC values for linear and quadratic Poisson point process
models for $\log(\lambda)$ of \textup{Angophora costata} presence. Model
fitted at the 1 km by 1 km resolution}\label{tab:AIC}
\begin{tabular*}{275pt}{@{\extracolsep{\fill}}lc@{}}
\hline
\textbf{Model} & \textbf{AIC}\\
\hline
Linear terms only & 5363.6\\
Quadratic (additive terms only) & 4763.4\\
Quadratic (interactions included) & 4400.6\\
\hline
\end{tabular*}
\end{table}

\begin{figure}[b]

\includegraphics{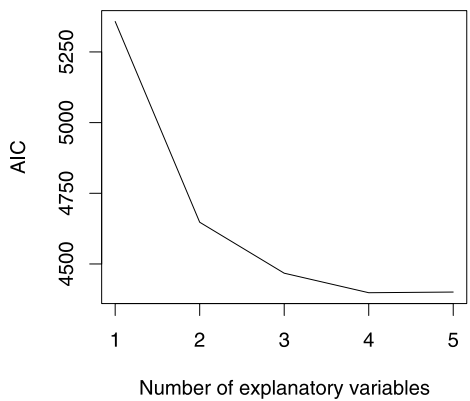}

\caption{Results of all-subsets selection, expressed as AIC of the
best-fitting model at each level of complexity. The respective
best-fitting models included minimum temperature, then minimum and
maximum temperature, then annual rainfall was added, then fire count,
and finally wetness. The best-fitting model included four explanatory
variables (all variables except wetness).}\label{fig:AIC}
\end{figure}

All-subsets selection considered a total of 32 models, and found that
the best-fitting model included four variables (Figure~\ref{fig:AIC})---all except for ``wetness.'' Parameter estimates and standard errors
for this best-fitting model are given in Table~\ref{tab:params}(a).

%f3 ###

An image of the fitted intensity surface from the best-fitting model is
presented in Figure~\ref{f:angoPoints}(c). The regions of highest
predicted intensity are near the coast and just north of Sydney, which
are indeed where the highest density of presence points appeared in
Figure~\ref{f:angoPoints}(a). We also compared intensity surfaces fitted
at different spatial resolutions, and note that they appear identical
when quadrature points are selected in a $500\times500$ m, $1\times
1$ km, or $2\times2$ km grid, and that irrespective of spatial
resolution, regions of higher intensity had 0.05--0.2 expected \textit{A.\
costata} records per square kilometer, as in Figure~\ref
{f:angoPoints}(c). Note this is in contrast to logistic regression, where
fitted probabilities in any given location are a function of number of
pseudo-absences, and vary by a factor of 16 when moving from a $2\times
2$ km to a $500\times500$ m grid.

\begin{figure}

\includegraphics{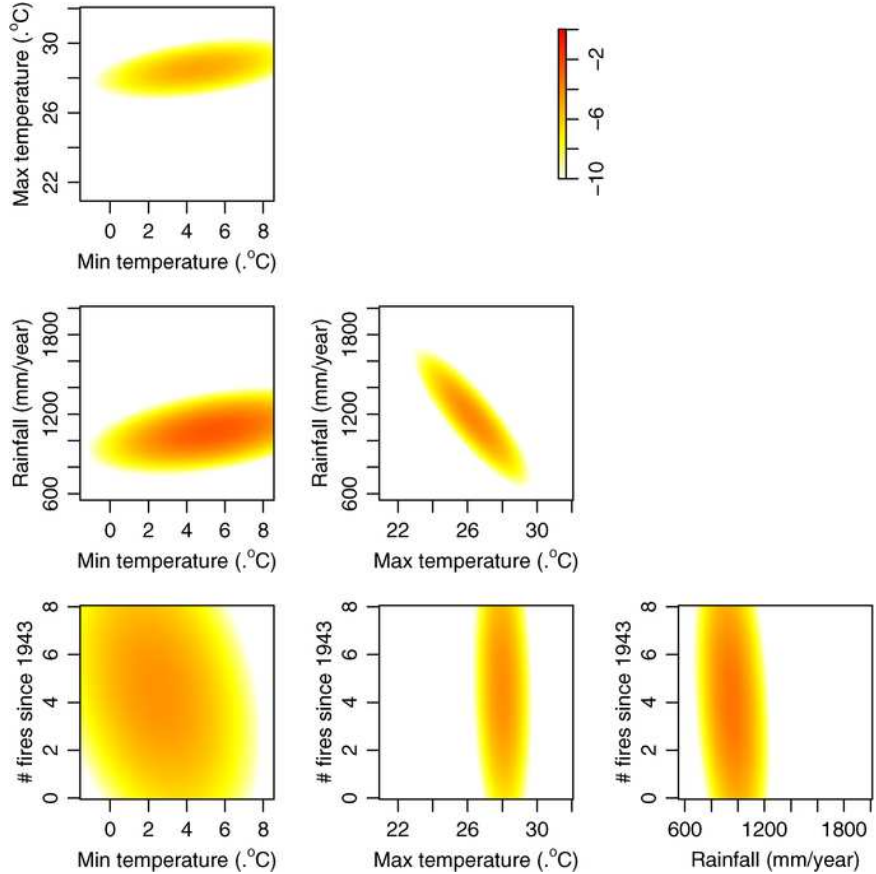}

\caption{Image plots of the joint effects of environmental variables on
predicted log (intensity) of \textup{A.\ costata} records. Darker areas
of the image correspond to higher predicted values of $\log(\lambda
_i)$. Note the strong and highly correlated response of intensity to
maximum temperature and annual rainfall, and the relatively weak
response to \# fires.}\label{f:contour}
\end{figure}

To assist in interpreting parameters from the best-fitting model, we
have constructed image plots of the fitted intensity in ``environmental
space'' to elucidate the nature of the effect of each environmental
variable on intensity of \textit{A.\ costata} records (Figure~\ref
{f:contour}). It can be seen in Figure~\ref{f:contour} that there is a
strong and negatively correlated response to maximum temperature and
annual rainfall. Of the four environmental variables, the response to
number of fires appears to be the weakest, with little apparent change
in predicted intensity as number of fires increased, and no observable
interaction with the three climatic variables.

%t2 ###
\begin{table}
\tabcolsep=0pt
\caption{Parameter estimates and their standard errors for \textup{(a)} the
Poisson point process model with a $1\times1$~km regular grid of
quadrature points; \textup{(b)} The logistic regression model with a $1\times
1$ km regular grid of pseudo-absences; \textup{(c)} The logistic model with 1000
randomly chosen pseudo-absence points. In each case we fitted a
quadratic model of minimum temperature (MNT), maximum temperature
(MXT), annual rainfall (RA), and fire count (FC). Notice that with few
exceptions, terms are equivalent to 2--3 significant figures for the
models fitted over a regular grid. But this is not the case for \textup{(c)},
and, in particular, standard errors are all 30--80\% larger}\label{tab:params}
\begin{tabular*}{\textwidth}{@{\extracolsep{\fill}}ld{5.5}d{3.5}d{5.5}d{3.5}d{5.6}d{3.5}@{}}
\hline
& \multicolumn{2}{c}{\textbf{(a)}} & \multicolumn{2}{c}{\textbf{(b)}} &\multicolumn{2}{c@{}}{\textbf{(c)}}\\[-5pt]
& \multicolumn{2}{c}{\hrulefill}& \multicolumn{2}{c}{\hrulefill}& \multicolumn{2}{c@{}}{\hrulefill}\\
\textbf{Term} & \multicolumn{1}{c}{$\bolds{\hat{\beta}_j}$} & \multicolumn{1}{c}{$\bolds{\mathit{se}(\hat
{\beta}_j)}$} & \multicolumn{1}{c}{$\bolds{\hat{\beta}_j}$} & \multicolumn{1}{c}{$\bolds{\mathit{se}(\hat{\beta}_j)}$} &
\multicolumn{1}{c}{$\bolds{\hat{\beta}_j}$} & \multicolumn{1}{c@{}}{$\bolds{\mathit{se}(\hat{\beta}_j)}$}\\
\hline
Intercept & -2130 & 169.4 & -2119 & 171 &  -1999 & 227 \\
MNT & -16.3 & 3.0 & -16.2 & 3.0 & -9.91 & 4.2 \\
MNT$^2$ & -0.21 & 0.027 & -0.205 & 0.028 & -0.185 & 0.050 \\
MXT & 128.7 & 10.1 & 128.1 & 10.1 & 120.2 & 13  \\
MNT${}*{}$MXT & 0.539 & 0.090 & 0.535 & 0.091 & 0.377 & 0.13\\
MXT$^2$ & -1.98 & 0.15 & -1.97 & 0.15 & -1.84 & 0.20 \\
RA & 0.759 & 0.065 & 0.755 & 0.066 & 0.714 & 0.089 \\
MNT${}*{}$RA & 0.00345 & 0.00065 & 0.00339 & 0.00065& 0.00147 & 0.00096 \\
MXT${}*{}$RA & -0.0218 & 0.0019 & -0.0216 & 0.0019 & -0.0203 & 0.0025 \\
RA$^2$/1000 & -0.0819 & 0.0072 & -0.0815 & 0.0072
& -0.0749 & 0.010 \\
FC & 6.24 & 3.37 & 5.98 & 3.42 & 4.08 & 4.9 \\
MNT${}*{}$FC & -0.101 & 0.040 & -0.101 & 0.041 &-0.207 & 0.070 \\
MXT${}*{}$FC & -0.123 & 0.10 & -0.115 & 0.010 &
-0.0601 & 0.15 \\
RA${}*{}$FC & -0.00174& 0.00066 & -0.00171 & 0.00067&
-0.000952 & 0.00095 \\
FC$^2$ & -0.127 & 0.024 & -0.123 & 0.024 & -0.107
& 0.041 \\
\hline
\end{tabular*}
\end{table}
%f4 ###

For the purpose of comparison, parameter estimates and their standard
errors are reported not just for the point-process model fit, but also
for the analogous logistic regression model in Table~\ref{tab:params}(b),
for a model fitted with quadrature points sampled in a $1\times1$ km
regular grid. Note that most parameter estimates and standard errors
differ by less than 1\% between the logistic regression and point
process models, as expected given Theorems~\ref{t:ll}--\ref{t:weights}.

We also report results when logistic regression is applied to
$m-n=1000$ pseudo-absences at randomly selected locations [Table~\ref
{tab:params}(c)]. This is at the lower end of the range typically used
[\citet{PearceBoyce}; \citet{ElithLeathwick2007}; \citet{Hernandezetal}] for
pseudo-absence selection in ecology. Note that the standard errors are
substantially larger in this case, and no parameter estimates are
correct past the first significant figure. This result exemplifies how
current practice in ecology regarding the number of pseudo-absences
($m-n$) can lead to poor results. Instead, it is advisable to consider
the sensitivity of results to different choices of $m-n$, along the
lines of Figure~\ref{f:res}.

\begin{figure}

\includegraphics{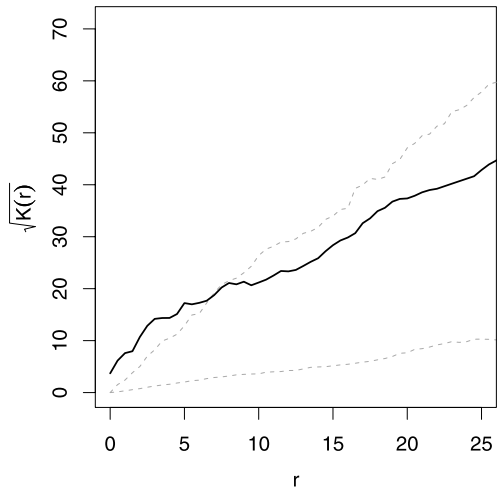}

\caption{Goodness of fit plot of the quadratic Poisson process
model---inhomogeneous $K$-function (solid line) with simulation envelope
(broken lines). The envelope gives 95\% confidence bands as estimated
from 500 simulated data sets. Note that the $K$-function falls within
those bounds over most of its range, although with a possible departure
for $r<6$ km.}\label{f:Kfn}
\end{figure}

To explore the goodness of fit of the best-fitting model, an
inhomogeneous $K$-function [\citet{BaddeleyKinhom}] was plotted using
the \texttt{kinhom} function from the \texttt{spatstat} package on
\texttt{R} [\citet{Spatstat}], and simulation envelopes around the
fitted model were constructed. See \citet{DiggleBook} for details
concerning the use of $K$-functions to explore goodness of fit of point
process models. The inhomogeneous $K$-function, as its name suggests,
is a generalization of the $K$-function to the nonstationary case. It
is defined over the region $\mathcal{A}$ as
\[
K_{\mathrm{inhom}}(r) = \frac{1}{|\mathcal{A}|} E \biggl\{ \sum_{y_i\in\mathbf
{y}}\sum_{y_j\in\mathbf{y}\backslash\{y_i\}} \frac{I(\|
y_i-y_j \|<r)}{\lambda_i\lambda_j} \biggr\}.
\]
$K_{\mathrm{inhom}}$ reduces to the usual $K$ function for a stationary
process, and like the usual $K$-function, can be used to diagnose
whether there are interactions in the point pattern $\mathbf{y}$ [\citet
{BaddeleyKinhom}]. Results (Figure~\ref{f:Kfn}) suggest a reasonable
fit of this model to the data, although with some lack of fit at small
spatial scales ($r<6$ km) suggestive of possible clustering in the data.
One possible method of modeling this clustering is to fit an
area-interaction process [\citet{BaddeleyAreaInter}], using the \texttt
{spatstat} package. We have repeated analyses using such a model and
found results to be generally consistent with those presented here,
except of course that the equivalence with logistic regression no
longer holds.

%f5 ###

%s5 ###
\section{Discussion}\label{s:discuss}
In this paper we have proposed the use of Poisson point process models
for the analysis of presence-only data in ecology, an important and
widely-studied problem to which this methodology is well suited. We
have also shown that this method is approximately equivalent to
logistic regression, when a suitable number of regularly or randomly
spaced pseudo-absences are used, hence, we provide a link between the
proposed method and the approach most commonly used in ecology at the
moment. But this raises the question: why use point process models, if
the method currently being used is (asymptotically) equivalent to
logistic regression anyway? Several reasons are listed below.

Recall that in Section~\ref{s:back} we argued that the pseudo-absence
approach has problems with model specification, interpretation, and
implementation. We argue that each of these difficulties is resolved by
using a point process modeling framework. \textit{Model specification}---we believe that a point process model as in Section~\ref{s:ppm} is a
plausible model for the data generation mechanism for presence-only
data. In contrast, the logistic regression approach involves generating
new data in order to fit a model originally designed for a different
problem (analysis of binary data not analysis of point-events). Hence,
the pseudo-absence approach as it is usually applied appears to involve
coercing the data to fit the model rather than choosing a model that
fits the original data. \textit{Interpretation}---in the logistic
regression approach we model $p_i$, the probability that a given point
event is a presence not a pseudo-absence. This quantity has no physical
meaning and clearly its interpretation is sensitive to our method of
choice of pseudo-absences (and typically each $p_i\rightarrow0$ as
$m\rightarrow\infty$). In contrast, the intensity at a point $\lambda
_i$ has a natural interpretation as the (limiting) expected number of
presences per unit area, and will \textit{not} be sensitive to choice of
quadrature points, provided that the number of quadrature points is
sufficiently large. \textit{Implementation}---in Section~\ref{s:ppm} we
explain that point process models offer a framework for choosing
quadrature points. Specifically, equation~(\ref{e:ppmLL}) is used to
estimate the point process log-likelihood, and progressively more
quadrature points are added until convergence of $l_{\mathrm{ppm}}(\hat
{\beta};\mathbf{y},\mathbf{y}_0,\mathbf{w})$ is achieved as in
Figure~\ref{f:res}(a). No such framework for choice of pseudo-absences is
offered by logistic regression, and instead choice of the location and
number of pseudo-absences is ad hoc, with potentially poor
results [Table~\ref{tab:params}(c)]. Ecologists are concerned about the
issues of how many pseudo-absences to choose [\citet{PearceBoyce}],
where to put them [\citet{ElithLeathwick2007}; \citet{Zarnetskeetal}; \citet{Phillipsetal09}], and what spatial
resolution to use in model-fitting [\citet{Guisanetal2007Grain}; \citet{ElithLeathwickSDM}], all issues that have natural
solutions given a point process model specification of the problem, as
in Section~\ref{s:ppm}.

It should be emphasized that we have demonstrated equivalence of point
process modeling and pseudo-absence logistic regression only for large
numbers of pseudo-absences and only for pseudo-absences that are either
regularly spaced or located uniformly at random over $\mathcal{A}$.
Current practice concerning selection of pseudo-absences in the ecology
literature does not always involve sampling at random over $\mathcal
{A}$ [e.g., \citet{Hernandezetal}] and does not involve sampling
sufficiently many pseudo-absences for model convergence. Instead,
choice of the number of pseudo-absences is ad hoc, and a total
of 1000--10,000 pseudo-absences is usually used
[\citet{ElithLeathwick2007}; \citet{Hernandezetal}], although sometimes even fewer
[\citet{Zarnetskeetal}]. On Figure~\ref{f:res}, 1000--10,000 corresponds
to a resolution of about 4--8 km, for which model convergence has not
been achieved. When fitting a model using just 1000 pseudo absences,
some parameter estimates are not equivalent to the high-resolution fits
to even one significant figure, and all standard error estimates were
larger by 30--80\% (Table~\ref{tab:params}).

While only Poisson point processes were considered in this paper, the
methodology implemented in Section~\ref{s:data} can be generalized to
incorporate interactions between points in a straightforward fashion
[\citet{Spatstat}]. However, the links between point process models and
logistic regression identified in Section~\ref{s:pseudo} may be lost in
this more general setting.

One issue not touched on in this paper is the problem of observer bias---that the likelihood of a species being reported is a function of
additional variables related to properties of the observer and not of
the target species, such as variation in the level of accessibility of
different parts of the region $\mathcal{A}$. For example, the high
number of \textit{A.\ costata} records just north of Sydney may be due in
part to proximity to a large city, rather than simply being due to
environmental conditions being suitable for \textit{A.\ costata}. This
issue will be addressed in a related article.

\begin{appendix}
%s6 ###
\section{Proof of theorems}\label{s:appendix}

%s6.1 ###
\subsection{\texorpdfstring{Proof of Theorem~\protect\ref{t:ll}}{Proof of Theorem 3.1}}

The proof involves two steps. The first step involves showing that
$l_{\mathrm{bin}}(\cdot)$, as a function of $p_i$, is asymptotically
equivalent to $l_{\mathrm{ppm}}(\cdot)$ when written as a function of
$\lambda_i$. The second step involves showing that given the
definitions of $p_i$ and $\lambda_i$ in equations (\ref{e:ppmModel})
and (\ref{e:binModel}), we can replace one with the other without
affecting the order of approximation.

Specifically, the log-likelihood function for $U$ can be written as
\[
l_{\mathrm{bin}}(\gamma; \mathbf{y}, \mathbf{y}_0 ) = \sum_{i=1}^n \log(p_i) + \sum_{i=n+1}^m \log(1-p_i)
\]
and a Taylor expansion of $\log(1-p_i)$ yields
\[
= \sum_{i=1}^n \log(p_i) + \sum_{i=n+1}^m  \{ p_i + O(p_i^2)  \},
\]
but it can be seen from equation~(\ref{e:Bayes}) that $p_i=O(m^{-1})$ and, hence, $\sum_{i=1}^n p_i = O(m^{-1})$ for fixed $n$, so
%e6.1 ###
\begin{eqnarray}\label{e:ppmForm}
l_{\mathrm{bin}}(\gamma; \mathbf{y}, \mathbf{y}_0 ) &=& \sum_{i=1}^n \log(p_i) + \sum_{i=n+1}^m p_i + O(m^{-1}) \nonumber\\
&=& \sum_{i=1}^n \log(p_i) + \sum_{i=1}^m p_i + O(m^{-1}) \\
&=& \sum_{i=1}^m  \{ I(i\in \{1,\ldots, n\}) \log(p_i) - p_i  \} + O(m^{-1}).\nonumber
\end{eqnarray}
Note that equation~(\ref{e:ppmForm}) has the form of the Poisson point
process log-likelihood, but with all weights set to one and $p_i$ being
used in place of $\lambda_i$. We will now derive a relation between
$p_i$ and $\lambda_i$ which motivates the replacement of $p_i$ by
$\lambda_i$.

First note that the Taylor expansion of $\log(1-x)$ implies both that\vspace*{1pt}
$\log(1-p_i) = O(m^{-1})$, and that $\log(m-n)= \log(m) + \log(1-n/m) =
\log(m) + O(m^{-1})$. So from equation~(\ref{e:binModel}),
\begin{eqnarray*}
\log p_i &=& \gamma_0 - \log(m-n) + \sum_{j=1}^k x_{ij}\gamma_j - \log(1-p_i)\\
&=& \gamma_0 - \log(m) + \sum_{j=1}^k x_{ij}\gamma_j + O(m^{-1}).
\end{eqnarray*}
This has the form of equation~(\ref{e:ppmModel}), where $\beta=\gamma -
J_m$. So when $\beta=\gamma - J_m$, $\log p_i = \log\lambda_i +
O(m^{-1})$, and $\sum_{i=1}^m p_i = \sum_{i=1}^m \lambda_i
\{1+O(m^{-1}\} = \sum_{i=1}^m \lambda_i + O(m^{-1})$. Now plugging
these results into equation~(\ref{e:ppmForm}) yields
$l_{\mathrm{ppm}}(\gamma-J_m; \mathbf{y}, \mathbf{y}_0, \mathbf{1}) +
O(m^{-1})$, completing the proof.\qed

%s6.2 ###
\subsection{\texorpdfstring{Proof of
Theorem~\protect\ref{t:weights}}{Proof of Theorem 3.2}}

The proof follows by inspection of the score equations. Specifically,
let $s_j(\beta;\mathbf{w}) = \frac{\partial}{\partial
\beta_j}l_{\mathrm{ppm}}(\beta;\mathbf{y},\mathbf{y}_0,\mathbf{w})$. From
equation~(\ref{e:ppmLL}), for $j\in\{1,\ldots,k\}$,
%e6.2 ###
\begin{equation}\label{e:score}
s_j(\beta;\mathbf{w}) = \sum_{i=1}^m x_{ij}\lambda_i w_i \biggl( \frac{z_i-\lambda_i}{\lambda_i}  \biggr) = \sum_{i=1}^m x_{ij} w_i(z_i-\lambda_i),
\end{equation}
where $z_i = \frac{I(i\in {1,\ldots,n})}{w_i}$. If $j=0$, equation~(\ref{e:score}) holds but with $x_{ij}=1$ for each $i$.

Now $\hat{\beta}$ satisfies $s_j(\hat{\beta};\frac{|\mathcal{A}|}{m}\mathbf{1}) = 0$ for each $j$, that is,
%e6.3 ###
\begin{equation}\label{e:scoreGrid}
0 = \sum_{i=1}^m x_{ij} \biggl( I(i\in {1,\ldots,n})- \hat{\lambda}_i \frac{|\mathcal{A}|}{m} \biggr),
\end{equation}
where from equation~(\ref{e:ppmModel}), $\log(\hat{\lambda}_i) = \hat{\beta}_0 + \sum_{j=1}^k x_{ij}\hat{\beta}_1$.

$\hat{\gamma}$ satisfies $s_j(\hat{\gamma}-J_m;\mathbf{1}) = 0$, for each $j$,
%e6.4 ###
\begin{equation}\label{e:scoreI}
0 = \sum_{i=1}^m x_{ij} \bigl( I(i\in {1,\ldots,n})- \tilde{\lambda}_i  \bigr),
\end{equation}
where $\tilde{\lambda}_i$ is the maximum likelihood estimator of
$\lambda_i$ for
$l_{\mathrm{ppm}}(\gamma-J_m;\mathbf{y},\mathbf{y}_0,\mathbf{1})$,
which satisfies $\log(\tilde{\lambda}_i) = \hat{\gamma}_0 -\log m +
\sum_{j=1}^k x_{ij}\hat{\gamma}_1$.

The solutions to equations~(\ref{e:scoreGrid}) and (\ref{e:scoreI}) are
related by the identity
$\tilde{\lambda}_i=\hat{\lambda}_i\frac{|\mathcal{A}|}{m}$ for each
$i$, and if we take the logarithm of both sides,
\[ \hat{\gamma}_0 -\log m + \sum_{j=1}^k x_{ij}\hat{\gamma}_j = \hat{\beta}_0 + \sum_{j=1}^k x_{ij}\hat{\beta}_j  +\log|\mathcal{A}| -\log m. \]
Provided that the design matrix $X$ has full rank,
$\hat{\gamma} = \hat{\beta} + J_{|\mathcal{A}|}$.

Also, note that the $(j,j')$th element of the Fisher information matrix of $l_{\mathrm{ppm}}(\beta;\mathbf{y},\mathbf{y}_0,\mathbf{w})$ is
%e6.5 ###
\begin{equation}\label{e:Fisher}
I_{jj'}(\beta;\mathbf{w}) =  -E \biggl( \frac{\partial^2}{\partial \beta_j\beta_{j'}} l_{\mathrm{ppm}}(\beta;\mathbf{y},\mathbf{y}_0,\mathbf{w})  \biggr)  = \sum_{i=1}^m x_{ij} w_i x_{ij'} \lambda_i
\end{equation}
and so $I_{jj'}(\hat{\beta};\frac{|\mathcal{A}|}{m}\mathbf{1}) = \sum_{i=1}^m x_{ij} x_{ij'} \frac{|\mathcal{A}|}{m} \hat{\lambda}_i = \sum_{i=1}^m x_{ij} x_{ij'} \tilde{\lambda}_i = I_{jj'}(\hat{\gamma}-J_{m}; \mathbf{1})$ for each $(j,j')$. This completes the
proof.\qed

%s6.3 ###
\subsection{\texorpdfstring{Proof of Theorem~\protect\ref{t:random}}{Proof of Theorem 3.3}}

Let $\delta = \hat{\gamma}-\hat{\beta}-J_{|\mathcal{A}|}$. We will
prove the theorem by using a Taylor expansion of the score equations
for $l_{\mathrm{ppm}}(\beta;\mathbf{y},\mathbf{y}_0,\mathbf{1})$ to show that
for fixed $n$ and $m\rightarrow\infty$, $\delta
\stackrel{\mathcal{P}}{\rightarrow} 0$.

Let $\mathbf{S}(\beta; \mathbf{1})$ be the vector of score equations
whose $j$th element is $s_j(\beta;
\mathbf{1})=\frac{\partial}{\partial\beta_j}l_{\mathrm{ppm}}(\beta;\mathbf{y},\mathbf{y}_0,\mathbf{1})
$ and let $\mathbf{I}(\beta; \mathbf{1})$ be the corresponding\vspace*{-1pt} Fisher
information matrix. A Taylor expansion of $\mathbf{S}(\hat{\gamma}-J_m;
\mathbf{1})$ about $\mathbf{S}(\hat{\beta}+J_{|\mathcal{A}|/m};
\mathbf{1})$ yields
%e6.6 ###
\begin{equation}\label{e:Taylor}
\mathbf{S}(\hat{\gamma}-J_m; \mathbf{1}) =
\mathbf{S}\bigl(\hat{\beta}+J_{|\mathcal{A}|/m}; \mathbf{1}\bigr) -
\mathbf{I}\bigl(\hat{\beta}+J_{|\mathcal{A}|/m}; \mathbf{1}\bigr)\delta
+ O_p(\|\delta\|^2).
\end{equation}
The left-hand side is zero, because it is evaluated at the maximizer of
$l_{\mathrm{ppm}}(\beta;\mathbf{y},\mathbf{y}_0,\mathbf{1})$. Also, evaluating
$\lambda_i$ at $\hat{\beta}+J_{|\mathcal{A}|/m}$ gives $\hat{\lambda}_i
\frac{|\mathcal{A}|}{m}$, and substituting this into
equation~(\ref{e:score}) at $\mathbf{w}=\mathbf{1}$,
\begin{eqnarray*}
s_j\bigl(\hat{\beta}+J_{|\mathcal{A}|/m}; \mathbf{1}\bigr) &=& \sum_{i=1}^n x_{ij} - \sum_{i=1}^m x_{ij} \hat{\lambda}_i \frac{|\mathcal{A}|}{m}\\
&\stackrel{\mathcal{P}}{\rightarrow}& \int_{y\in \mathcal{A}} x_j(y) \hat{\lambda}(y) \,dy
\end{eqnarray*}
from the weak law of large numbers. But this is the derivative of
$l(\beta;\mathbf{y})$ from equation~(\ref{e:ppmint}), evaluated at the
maximum likelihood estimate $\hat{\beta}$, and so it equals zero for
each $j$ and, hence, $\mathbf{S}(\hat{\beta}+J_{|\mathcal{A}|/m};
\mathbf{1}) \stackrel{\mathcal{P}}{\rightarrow} \mathbf{0}$. Similarly,
for each $(j,j')$, $I_{jj'}(\hat{\beta}+J_{|\mathcal{A}|/m};
\mathbf{1}) \stackrel{\mathcal{P}}{\rightarrow} I_{jj'}(\hat{\beta})$,
the $(j,j')$th element of the Fisher information matrix for
$\hat{\beta}$ from $l(\beta;\mathbf{y})$. So returning to
equation~(\ref{e:Taylor}),
\[
\delta = \mathbf{I}\bigl(\hat{\beta}+J_{|\mathcal{A}|/m}; \mathbf{1}\bigr)^{-1}
\mathbf{S}\bigl(\hat{\beta}+J_{|\mathcal{A}|/m}; \mathbf{1}\bigr) +
O_p(\|\delta\|^2) \stackrel{\mathcal{P}}{\rightarrow} 0,
\]
completing the proof.\qed
\end{appendix}

\section*{Acknowledgments}

Thanks to the NSW Department of Environment and Climate Change for
making the data available, and to Dan Ramp and Evan Webster at UNSW for
assistance processing the data and advice.
Thanks to David Nott, William Dunsmuir and Simon Barry for helpful discussions.

\printaddresses

\end{document}